\documentclass{ws-procs9x6-cpt16}

\begin{document}

\newcommand{\refeq}[1]{(\ref{#1})}
\def\etal {{\it et al.}}

\title{Searches for Exotic Transient Signals with a\\
Global Network of Optical Magnetometers for Exotic Physics}

\author{S.\ Pustelny}

\address{Institute of Physics, Jagiellonian University, 
30-383 Krak\'ow, Poland}

\author{On behalf of the GNOME Collaboration\footnote{The collaboration website: \url{budker.uni-mainz.de/gnome/}}}

\begin{abstract}
In this letter, we describe a novel scheme for searching for physics beyond the Standard Model. The idea is based on correlation of time-synchronized readouts of distant ($\gtrsim$100~km) optical magnetometers. Such an approach limits hard-to-identify local transient noise, providing the system with unique capabilities of identification of global transient events. Careful analysis of the signal can reveal the nature of the events (e.g., its nonmagnetic origin), which opens avenues for new class of exotic-physics searches (searches for global transient exotic spin couplings) and tests of yet unverified theoretical models. 
\end{abstract}

\bodymatter

\section{Global network of optical magnetometers}

Modern optical magnetometers\cite{Budker2013Optical} enable the detection of magnetic fields with a sensitivity reaching or even exceeding 1~fT/Hz$^{1/2}$. Yet, the magnetometers are not sensitive to magnetic field \textit{per se} but rather to its effect on energies of light-coupled magnetic sublevels. With this respect, optical magnetometers are intrinsically sensitive to any spin couplings, including exotic ones. This underlies the foundations for searches for physics beyond the Standard Model and formed the basis for many experiments over the last several decades (see, for example, Ref.~\refcite{Kimball2013Tests} and reference therein).

Despite the high sensitivities of the experiments searching for physics beyond the Standard Model, all the experiments have provided null results.\cite{Kimball2013Tests}  This has falsified several theoretical models, leading to their redefinition. However, to reach the precision, long-time averaging of the detected signals is required. This brings an important question: {\it what if the sought couplings (signals) are of time dependent (oscillatory or transient) character?} Obviously, in such a case, the averaging would lead to deterioration of the signal of interest rather than increase of a signal-to-noise ratio.

Over the last years, several theoretical models postulating the existence of time-dependent exotic spin couplings have been proposed. Such couplings would either lead to oscillating signals, e.g., due to oscillating electric dipole moment\cite{Hill2015Axion} or dark-matter couplings,\cite{Zhang2015Unconventional} or to transient signals, e.g., due to a passage through a topological defect (e.g., domain wall) of an exotic field\cite{Pospelov2013Detecting} or interaction with a jet of exotic particles.\cite{Arvanitaki2015Discovering} Detection of oscillating couplings requires new experiments, which are currently under construction in Mainz, Germany and Boston, USA (Cosmic Axion Spin Precession Experiment)\cite{Budker2013Cosmic} and Stanford, USA (Dark-matter radio).\cite{Chaudhuri2015Radio} Similarly, the transient signals will be investigated with a Global Network of Optical Magnetometers for Exotic physics (GNOME),\cite{Pustelny2013Global} described here.

GNOME is a network of synchronized optical magnetometers separated by a distance of hundreds or even thousands of kilometers. Each magnetometer is characterized with a femtotesla sensitivity and operates in a magnetically shielded, precisely controlled magnetic environment. In such a measurement, the magnetometer readouts are principally static, but the nature of the measurements burdens the signals with noise. The spectral characteristics of the noise may be quite complex, with frequency-independent (e.g., due to photon shot noise), $1/f$ (e.g., due to vibrations), and band (e.g., due to AC lines) components. Additionally, infrequent spikes (in the time domain) can be observed in the signals. The nature of this noise is particularly hard to identify and eliminate, making untriggered transient signals difficult to investigate.

Despite the complexity of the magnetometer readouts, the observed noise is typically local and hence the readouts of even same-type devices are characterized with different dependence. Particularly, it is unlikely to observe correlated transient signals in two or more magnetometers triggered by uncorrelated (local) events. In such a case, correlation measurements provide information about global transient disturbances. The source of these disturbances may be magnetic (due to the Earth-magnetic-field change and imperfections of magnetic shieldings) but such cases may be falsified with proper vetoing techniques (e.g., via correlating the signals with readouts of outside-of-the-shield sensors), leaving the system sensitive to global disturbances of nonmagnetic origins.

From the point of view of the network operation, precise timing of the signal acquisition is of crucial importance. In GNOME, the timing is provided by the global positioning system (GPS), which offers a time precision better than 100~ns. At the same time, the global spread of the magnetometers loosens requirements for sampling rate as even light-speed-propagating perturbations require several to several tens of milliseconds to transit between different stations. To meet these requirements, the GNOME collaboration developed a data-acquisition system, offering the ability to multi-channel data acquisition with up to 1~kS/s sampling rate and sub-microsecond time accuracy. The systems are now installed in all GNOME stations to acquire synchronously the magnetometer readouts.

From the point of view of foreseen measurements, correlation of signals with the same spectral characteristics is essential. To meet this requirement, the magnetometer signals need to be standardized. This is achieved by the so-called whitening procedure, which consists of normalizing the signals by an exponentially weighted signal history.\cite{Pustelny2013Global} This procedure produces a gaussian distributed, zero-mean and unity variance data stream, which can be reliably correlated (each magnetometer contributes at the same level to the correlation signals).

\section{GNOME trial run}

\begin{figure}[t]
\begin{center}
\includegraphics[width=\hsize]{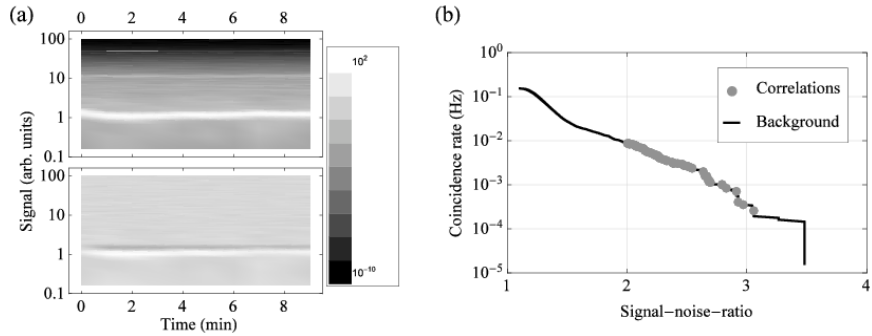}
\end{center}
\caption{(a) Spectrogram of 10-minute Krak\'ow magnetometer signal raw (top) and whitened (bottom) signal. (b) Coincidence rate between signals of Krak\'ow and Berkeley magnetometers.\label{fig:data}}
\end{figure}

The current GNOME consists of 5 stations situated in 4 countries (Germany, Poland, Switzerland, and the US).
Several other stations are under construction.
In the existing arrangement, the network performed a 24-hour trial run in March 2016. The signals from all magnetometers were stored and transmitted to the main server (Mainz, Germany), where they were processed. The spectrogram of a sample time series detected with the Krak\'ow magnetometer is shown in Fig.~\ref{fig:data}(a) (top panel). Similarly to signals of all other magnetometers (not shown), the data reveals strong $1/f$-noise but also two strong modulation components at 1~Hz (slowly drifting $\pm 0.05$~Hz) and 12~Hz (stable over time). As shown in bottom panel of Fig.~\ref{fig:data}(a), $1/f$-noise and the modulation components are strongly suppressed with the whitening procedure; the whole noise amplitude span was reduced by about 5 orders of magnitude leaving a detectable, yet strongly reduced trace only at 1~Hz. 

Whitening procedure enables the identification of transient events, manifesting as local maxima in the time-frequency spectrogram. However, from the point of view of GNOME operation, correlation measurements are of crucial importance. Thus, the correlations of two hour-long time series of Krak\'ow and Berkeley magnetometers [Fig.~\ref{fig:data}(b)] are presented. To perform reliable analysis, we first estimated the background by correlating time series separated by several tens of minutes, where there are no correlations. This enables identification of a false positive-event rate (solid line). Then, we correlated the signals with no time offset, i.e., a specific case for which the correlations could be observed (gray points). For such a case, several tens of events were identified, all consistent with the false-positive background. This result presents the ability to identify correlation between the stations but also demonstrates the necessity of improving performance of the magnetometers and developing more sophisticated (multi-station) correlation algorithms.

\section{Perspective of the network and conclusions}

The results of the trail run demonstrated the ability to identify transient events in signals of two magnetometers. Currently, the network magnetometers are being optimized to reach sensitivity of roughly 10~fT/Hz$^{1/2}$ and bandwidth of $>100$~Hz and 24/7 operation. The start of a multi-month run is planned at the end of 2016. This will enable a search for axion-like-field domain structure, exotic field bursts, and exotic coherent oscillations. However, the universality of the network makes it a unique tool for testing theories beyond the Standard Model.\footnote{The authors welcome suggestions regarding specific theories testable with GNOME and encourage potential collaborators to contact the network representatives.}

\section*{Acknowledgments}
The author acknowledges support from the Polish Ministry of Science and Higher Education within the Iuventus Plus program (grant 0390/IP3/2015/73).


\begin{thebibliography}{xx}

\bibitem{Budker2013Optical}
D. Budker and M. V. Romalis,
{\it Optical magnetometry}
Nature\ Phys.\ {\bf 3}, 227-234 (2007).

\bibitem{Kimball2013Tests}
D. F. Jackson Kimball and S. K. Lamoreux and T. E. Chupp
{\it Tests of fundamental physics with optical magnetometers}
in {\it Optical magnetometry}, D. Budker and D. F. Jackson Kimball Eds.
Cambridge University Press, Cambridge, 2013.

\bibitem{Hill2015Axion}
C. T. Hill
{\it Axion induced oscillating electric dipole moments}
Phys.\ Rev.\ D {\bf 91}, 111702 (2015).

\bibitem{Zhang2015Unconventional}
Z. Zhang ajd Y. Zhao
{\it Unconventional dark matter models: a brief review}
Sci. Bull. {\bf 60}, 986 (2015).

\bibitem{Pospelov2013Detecting}
M. Pospelov {\it et al.},
{\it Detecting Domain Walls of Axionlike Models Using Terrestrial Experiment}
Phys.\ Rev.\ Lett.\ {\bf 110}, 021803 (2013).

\bibitem{Arvanitaki2015Discovering}
A. Arvanitaki, M. Baryakhtar, and X. Huang
{\it Discovering the QCD axion with black holes and gravitational waves}
Phys.\ Rev.\ D {\bf 91}, 084011 (2015).

\bibitem{Budker2013Cosmic}
D. Budker, P. W. Graham, M. Ledbetter, S. Rajendran, and A. O. Sushkov
{\it Cosmic Axion Spin Precession Experiment (CASPEr)}
Phys.\ Rev. X {\bf 4}, 021030 (2014).
 
\bibitem{Chaudhuri2015Radio}
S. Chaudhuri {\it et al.},
{\it Radio for hidden-photon dark matter detection}
Phys.\ Rev.\ D {\bf 92}, 075012 (2015).

\bibitem{Pustelny2013Global}
S. Pustelny {\it et al.},
{\it The Global Network of Optical Magnetometers for Exotic physics (GNOME): A novel scheme to search for physics beyond 
the Standard Model}
Ann.\ Phys.\ {\bf 525}, 659-670 (2013).

\end{thebibliography}
\end{document}